\documentclass[a4paper,11pt]{article}
\usepackage{jinstpub} 
\usepackage{lineno}
\usepackage{graphicx}
\usepackage{cleveref} 
\usepackage[dvipsnames]{xcolor}
\usepackage{textcomp,nicefrac}
\usepackage{notoccite}
\usepackage{booktabs}
\usepackage{url}
\usepackage{etoolbox}

\title{\boldmath High-Resolution 3D-Printed Plastic Scintillators with Tertiary Dye}

\author[a]{C.J. Moore,}
\author[a,b]{M. Febbraro,}
\author[a,1]{J.J. Manfredi, \note{Corresponding authors.}}
\author[b]{A. Wood,}
\author[a]{D. Rutstrom,}
\author[b]{T. Ruland,}
\author[b]{B. Hackett,}
\author[b]{P.A. Hausladen}
\affiliation[a]{Department of Engineering Physics, Air Force Institute of Technology, 2950 Hobson Way, Wright-Patterson AFB, OH 45433, U.S.A}
\affiliation[b]{Oak Ridge National Laboratory TN, 37830 U.S.A}

\emailAdd{juan.manfredi@us.af.mil}

\abstract{
Additive manufacturing offers efficient production of plastic scintillators with nontrivial geometries using vat polymerization, allowing fabrication of geometries which would be difficult or even impossible to produce using conventional subtractive manufacturing. This work presents a novel photocurable scintillator formula that includes coumarin 450 as a tertiary dye to enable high-resolution 3D printing via the manipulation of the 405 nm cure light. Bulk photocured and 3D printed (with and without tertiary dye) samples were compared through observational assessment and spectral response. All samples showed pulse shape discrimination between neutron and gamma events.  Inclusion of the tertiary dye has minimal impact on emission spectrum and light output, but significant impact on print resolution as shown by comparison of printed high-complexity geometries and feature resolution test objects. With the use of a cure-limiting dye, unsupported features—such as freestanding pillars—were resolvable down to 0.7 mm. Even finer resolution at or below 0.1 mm was achieved in fully supported, integrated structures printed with off-the-shelf 405 nm desktop 3D printer. Scintillators demonstrated a light output up to 50 \% of EJ-200 with a PSD figure of merit up to 1.35 at 0.9 – 1.1 MeVee.}


\begin{document}
\date{}
\maketitle

\flushbottom

\section{Introduction}
\label{sec:introduction}
Organic scintillators are commonly used for detecting broad ranges of ionizing radiation due to their low cost, ruggedness, and ease of manufacture in large geometries. Plastic organic scintillators have been in use since the early 1950s \cite{Schorr1950} and are made up of long, linked chains of hydrocarbon materials that make the plastic material robust and machinable. Most plastic scintillators use aromatic monomers like vinyltoluene or styrene as a base, serving both as the primary building block for the polymer matrix but also as the primary excitation point for incoming radiation \cite{Birks1970}. Plastic scintillators are traditionally manufactured by thermal polymerization, in which a liquid monomer-based resin is heated over a period of days or weeks to incite polymerization. While possible with precise and expensive machining tools, making scintillators with non-standard geometries using this approach is often impractical or even impossible. The advent of additive manufacturing (AM), in particular using light-based photopolymerization, offers a new potential path for the manufacture of these radiation detectors at much faster timescales and with arbitrary geometry \cite{Kim,365Chandler}. Scintillating materials printed at high resolution offer potential applications in microfluidics \cite{Glennon2022}, high energy physics \cite{Berns2022,Weber2024}, and rare-event physics \cite{Hackett2024}. Current literature examples of AM plastics with high print resolution and good scintillation performance either rely on a copolymerization scheme that can cause nonuniformities and a change in opacity due to unreacted monomer \cite{Kim, Anand2025} or use a non-standard 365 nm wavelength \cite{365Chandler}. The present work presents a scintillator resin formulation that prints at 405 nm, relies on a non-aromatic oligomer base and cross-linkers, and has pulse shape discrimination (PSD) to distinguish between fast neutron and gamma events.

Leveraging the advances of AM for plastic scintillators requires fast curing resins that make optically clear plastics: one approach is to use light-based polymerization. However, resins incorporating aromatic structures with less reactive photopolymerizable end groups often exhibit significantly reduced polymerization rates, leading to solidification times of multiple hours for even small volumes \cite{Lim2019, Frandsen2023, Dolezal2023}. Therefore scintillator resins based on standard aromatic monomers \cite{Kim, Anand2025} need significantly longer curing times and result in poorer resolution. Alternatively, nonaromatic methacrylate and acrylate compounds due to their electron-rich carbon–carbon double bond (C=C) in the $\alpha,\beta$-unsaturated ester structure are highly reactive to light in combination with a photoinitiator, leading to curing times on the order of tens of seconds for a thin layer \cite{Mishnayot2014, Son2018, Frandsen2023}. However, because these compounds are not aromatic, they do not contribute to the scintillation process. Frandsen, et al. showed that plastics with nonaromatic monomers that cure within minutes can still produce good scintillation performance, albeit worse than fully aromatic materials, by directly exciting the primary fluor \cite{Frandsen2023}. This performance is quantified by light output (LO) and the Figure-of-Merit (FoM) metric for PSD.

Generally, organic scintillators contain multiple scintillating ingredients including the base monomer, a primary fluor which absorbs energy from the base via radiationless transfer before then emitting photons, and a secondary fluor or wavelength shifter which absorbs the primary fluor light and shifts it to longer wavelengths, normally to better match photodetector quantum efficiency. Common primary fluors or primary dyes for organic scintillators include polyphenyl hydrocarbons, oxazoles, and oxadiazole aryls. One of the most common primary fluors is 2,5-diphenyloxazole (PPO). This fluor was used in the seminal works of Brooks and Zaitseva, et al. that showed PSD capability in plastic scintillators by increasing PPO concentration to upwards of 15-30\%, enabling increased triplet-triplet annihilation which drives delayed fluorescence \cite{BROOKS, Zaitseva2012}. In formulations with nonaromatic bases where the primary fluor is directly excited by ionization radiation, this component is one of the most important in determining the LO, PSD, and the overall quality of the scintillating properties. Typical secondary fluors include 1,4-bis(2-methylstyryl)benzene (bis-MSB) and POPOP, but Frandsen, et al. showed that bis-MSB absorbs enough of the 405 nm cure light to result in a short cure depth which limited photo bulk polymerization \cite{Frandsen2023}. Instead, they used exalite 416 which has reduced absorbance at 405 nm.

\section{Motivation}

The current process of additively manufacturing scintillators via photopolymerization suffers from a tradeoff between LO and curing speed. Work from Chandler et al. found it is possible to additively manufacture PSD-capable plastics using less-standard 365 nm wavelength light with layer curing times lower than 9 seconds \cite{365Chandler} but resulted in a suppressed light yield of no more than 31\% of a benchmark plastic scintillator EJ-200. Work that produced higher LO by utilizing aromatic compounds like vinyltoluene required slow co-polymerization techniques that requires over a minute per layer for curing and risks causing the scintillators to significantly opaque \cite{Kim, Anand2025}, making use in AM pathways less viable.

Furthermore, current AM of scintillators with standard cure wavelengths (e.g., 405 nm) \cite{Frandsen2023, Kim, Anand2025} is more accurately described as a 2D+ technique. While it is possible to fabricate 3D geometries, the method is constrained by the need to prevent the projected curing region of each layer from extending too far beyond the foot-print of the previous layer. If this overlap occurs, excess resin between the target layer and build plate cures unintentionally, leading to a dramatic loss in both vertical and horizontal print resolution. This issue is particularly problematic for complex geometries—such as those with internal voids or expanding cross-sections along the build direction. 

The root of this issue lies in the long mean free path (MFP) of 405 nm light in typical scintillator resin formulations. As a result, light can penetrate deep into the resin vat, polymerizing regions well beyond the intended exposure plane. This compromises the fidelity of fine features and imposes a design constraint on printable geometries. One strategy to mitigate this involves tailoring the resin’s optical properties through the modification of already introduced dyes or fluors. Although secondary scintillating fluors---such as bis-MSB---are often selected to shift emitted scintillation light toward regions of higher photodetector efficiency, they also exhibit sufficient absorption at 405 nm to limit the MFP of the curing light and reduce curing depth. However, this dual role creates a trade-off: concentrations suitable for effective depth control are often too low to enable efficient scintillation, whereas concentrations that optimize scintillation prevent full layer polymerization due to excessive light absorption \cite{Frandsen2023}. As a result, a compromise must be made between geometric resolution and scintillation performance.

To overcome this limitation, the present work investigates the introduction of a \textit{tertiary dye}—a dye added independently of the primary and secondary scintillation fluors. The tertiary dye’s role is to specifically modulate the penetration of 405 nm curing light, thereby improving print resolution without compromising scintillation efficiency. By decoupling to the extent possible the functions of depth curing control and scintillation, each dye’s concentration can be tuned independently: the tertiary dye for cure depth optimization and the secondary dopant solely for scintillation wavelength shifting. However, selecting an appropriate tertiary dye is nontrivial. Coumarin 30 was found to be highly effective at limiting 405 nm light penetration and improving print resolution, yet even at concentrations as low as 0.005 wt.\% it caused undesirable green coloration in the printed geometries, which degraded optical clarity and introduced unwanted absorption of scintillation light.

Instead, this study explores the use of coumarin 450 as a tertiary dye within a non-aromatic formula using industry-standard 405 nm curing light. Coumarin 450 absorbs incoming near-UV light at 405 nm responsible for initiating polymerization via the TPO photoinitiator and re-emits it at longer wavelengths that do not contribute to radical formation. This spectral modulation attenuates the curing light’s penetration depth, confining polymerization to targeted regions within each layer without imparting significant color to the printed material. Moreover, coumarin 450 is effective at very low concentrations, enabling a reduction in cure volume without disrupting the polymerization process—a limitation observed with higher concentrations of bis-MSB when used as a secondary fluor \cite{Frandsen2023}. This paper presents the formulation and preparation process for this three-dye scintillator, as well as other two-dye photocured formulations for comparison. The effect of the tertiary dye on feature resolution and solidification time is explored, as well as the scintillation response of the final samples. Comparisons are made between these photocured plastics in conjunction with commercially available plastics of the same geometry.

\section{Materials \& Methods}

The following ingredients were used as received from Sigma-Aldrich: 99\% purity PPO, technical grade isobornyl acrylate (IBOA) with 200 ppm monomethyl ether hydroquinone as inhibitor, 1,6-hexanediol dimethacrylate (HDDMA) containing 75.0 ppm hydroquinone as inhibitor, and 97\% purity diphenyl(2,4,6-trimethylbenzoyl)phosphine oxide (TPO). BR-541MB difunctional aliphatic polyether urethane methacrylate (BR-541MB) oligomer was used as received from Bomar. Additional fluorescent dyes including 7,7’-Di(4-anisyl)-9,9,9’,9’-tetrapropyl-2,2’-bi-9H-fluorene (exalite 416) and 7-(ethylamino)-4,6-dimethyl-2H,-1-benzopyran-2-one (coumarin 450) were purchased and used as received from Luxottica-Exciton.

\subsection{Scintillator Formulation}
\label{sec:prep}

\begin{table}[t]
\caption{BR-541MB based scintillator two- and three-dye formulations by Component Fraction}
\label{formula}
\resizebox{\columnwidth}{!}{%
\begin{tabular}{@{}cccccccc@{}}
\toprule
 & \multicolumn{7}{c}{Component Fraction in Formula (wt.\%)} \\ \midrule
Formula & BR-541MB & IBOA & HDDMA & PPO & Exalite 416 & TPO & Coumarin 450 \\ \midrule
Two-dye & 39.7 & 27.79 & 11.91 & 20 & 0.5 & 0.1 &  \\
Three-dye & 39.68 & 27.776 & 11.904 & 20 & 0.5 & 0.1 & 0.04 \\ \bottomrule
\end{tabular}%
}
\end{table}

Two formula types were manufactured and assessed. The first was a fast photocurable resin that utilized all listed chemicals apart from coumarin 450 with PPO as the primary dye and exalite 416 as the secondary dye (i.e. the two-dye formula). Each layer of this formula cures within 1 minute but has excess polymerization where features see repeated exposure and therefore reduced print resolution in these areas \cite{Moore}. The second formula was similar to the first except for the addition of coumarin 450 as a tertiary dye; this formula is correspondingly referred to as the three-dye formula. Full formulas by component wt.\% are shown in \Cref{formula}.

Liquid resins were mixed according to each formula in preparation for polymerization into a solid plastic. Manufacture of all liquid resin batches followed a process of combining PPO, exalite 416, and, if used, coumarin 450 (all solid powders) to a 300 ml amber glass container. These containers were chosen to limit the amount of ambient light that could pass into the resin and preemptively cause polymerization. A high (20 wt.\%) concentration of PPO was utilized as the primary dye to offset the lack of an aromatic polymer base and maintain PSD capability. Higher concentrations of PPO had been previously tested, but resins with above 30 wt.\% PPO had leaching issues in which the PPO precipitated out of the solution. The secondary dye exalite 416 was added in low concentration (0.5 wt.\%) as outlined in previous work \cite{Frandsen2023}. The tertiary dye in the three-dye formula was evaluated at a large variation of concentrations from 0.01 to 0.1 wt.\%. Following the combination of the solid powders, liquid HDDMA and IBOA were added in a ratio of 30:70, respectively. Solutions sonicated at 50 $^\circ C$ for 30 minutes in alternating degas and sonication mode in order to ensure full dissolution. After dissolution BR-541MB was measured into the scintillator solution amber glass container. The solution was then re-sonicated for an additional 30 minutes to mix the viscous oligomer into the solution. Upon completion, the mixture was air-cooled and 0.1 wt.\% of TPO was added. The container was returned to the sonication bath for a final set of 30 minutes of sonication. Small (~15 g) amounts of solution were transferred from the large batch containers to small 20 ml scintillation vials for bulk photocuring tests. All containers were placed under vacuum at 7 mbar overnight to remove molecular oxygen gas from the solution.

\subsection{Scintillator Preparation \& Printing}

\begin{figure}[ht!]
    \centering    
    \includegraphics[width=.95\linewidth]{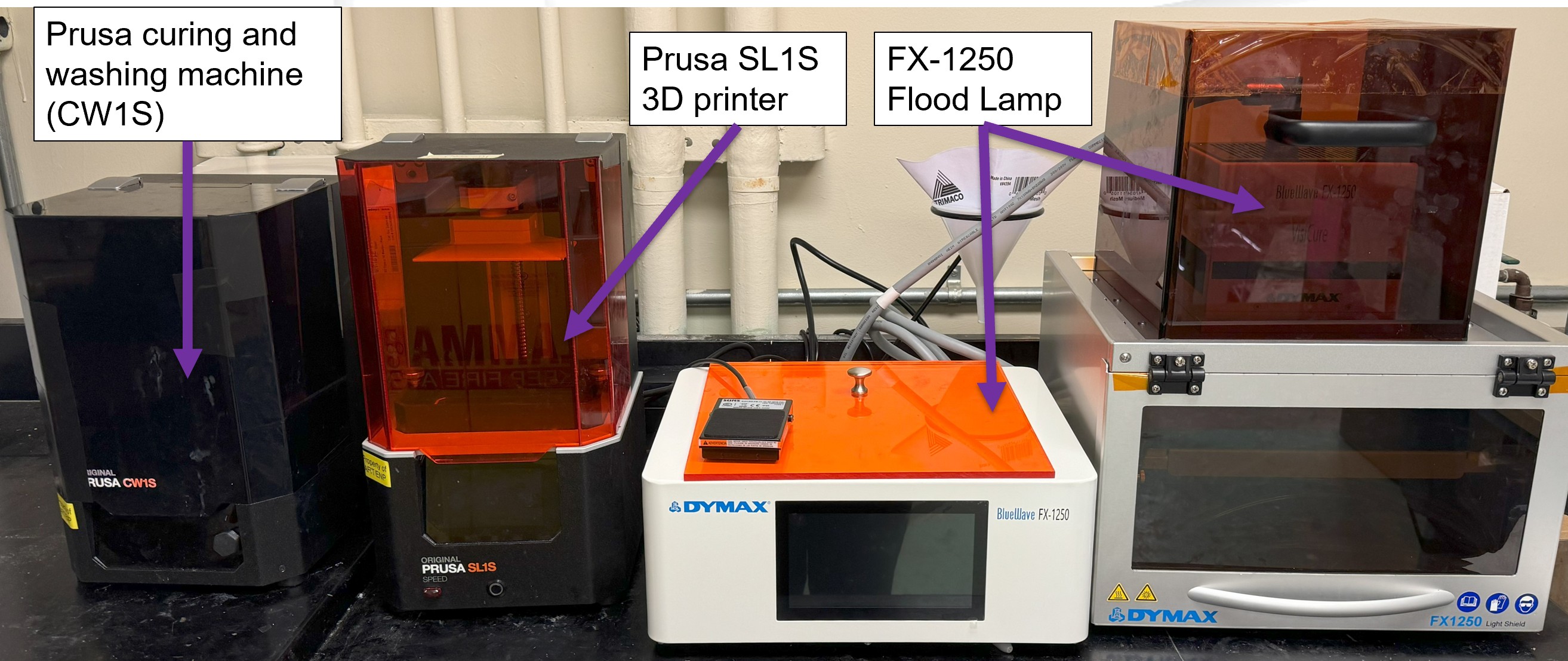}
    \caption{Printer assembly setup with vat polymerization SL1S 3D printer (center left), CW1S (far left), and post-curing FX-1250 lamp (center right and far right).}
    \label{Printersetup}
\end{figure}

Two separate photocuring protocols were followed to make plastic scintillators from the liquid resins. The first protocol involved ``bulk curing'' in which a volume of resin was photocured simultaneously while being bathed in constant 405 nm light \cite{Frandsen2023}.  The second protocol made use of a commercial 3D printer which cures resin one layer at a time.  Equipment for both protocols is shown in \Cref{Printersetup}.

The small scintillation vials containing resin solution were bulk cured by being exposed to 20 mW/cm$^2$ (on average) of 405 nm light within an Original Prusa Curing and Washing Machine (CW1S), first for one minute in increments of 10 seconds followed by an additional continuous 9 minutes. The short pauses in between 10-second exposures allow the formula to cool and prevent defects during the initial, highly exothermic polymerization. Samples were then exposed to 500 mW/cm$^2$ of 405 nm light within a Dymax BlueWave FX-1250 for 5 minutes to burn up any remaining TPO within the samples. Solid samples were broken from the glass vials, cut, sanded down to 1" in height, and the ends of the cylinders polished to remove imperfections down to 1 micron in size utilizing a semi-autonomous polishing machine. No three-dye formula samples were made using bulk photopolymerization protocol due to the limitation on the depth which the curing light could reach, causing the sample’s core to never fully solidify. 

3D printed scintillator geometries were prepared by pouring the amber container resin contents into a vat which sits within the Prusa SL1S 3D printer shown in \Cref{Printersetup}. This AM machine prints through a process called vat photopolymerization in which a liquid photopolymer held in a vat is selectively cured layer by layer by light-activated (typically near UV) polymerization. Digital light processing and stereolithography are alternative techniques that use this process, but Liquid Crystal Display (LCD), also known as masked stereolithography, is the most common example of vat polymerization for rapid, inexpensive 3D printing for the domestic ($\le\$$5000) market \cite{formlabsMSLALCD}. All geometries were printed in this work at 100 microns per layer and a curing time per layer between 20 and 45 seconds at the printer's designed 2.2 mW/cm$^2$ intensity \cite{NOVOKHATSKA2024} using the LCD vat polymerization technique. Upon print completion, samples were washed of uncured resin, post-cured for 10 minutes in the CW1S at an average of 20 mW/cm$^2$, and exposed to 5 minutes of 500 mW/cm$^2$ 405 nm light within a BlueWave FX-1250. The samples were finally cut to designed specifications and required minimal polishing of the ends of the printed geometry to remove imperfections down to 1 micron in size.

\subsection{Characterization \& Measurement}
Characterization of the scintillator formulas was performed through component fluor absorption and emission spectroscopy, bulk sample observational analysis, photoluminescence measurements, determination of minimum achievable feature resolution, and evaluation of scintillation performance.

Each photoactive component was assessed individually for its absorption on an Agilent Cary 7000 universal measurement spectrophotometer (UMS). The relevant compounds for testing were PPO, exalite 416, coumarin 450, and TPO. Absorption measurements were taken with 3.5 ml cuvettes and ethanol dilutions of 10$^{-5}$ to 10$^{-3}$ M depending on the intensity required over a range of 200 to 800 nm. Emission measurements for the same compounds utilized an Edinburgh FLS980 spectrofluorometer over a range of 200 – 1000 nm.

Each photopolymerized sample was first evaluated with respect to physical characteristics including relative discoloration and visual inspection for print defects and abnormalities. In instances where significant agitation or time occurred between degassing under vacuum and curing, air would be introduced back into solution. This causes features like bubbles to appear because air dissolved in solution can become trapped and join during solidification, forming a gap when the resin was photocured or 3D printed. Alternately, if too significant of a temperature gradient was present during curing, sections of the sample would polymerize, expand, and contract at different rates. This caused internal stresses that cause fractures and fissures in the final solidified sample. These non-representative features negatively impacted final LO through reflections and scattering. and therefore indicated problems with an initial methodology, leading to minimization of time between degassing and curing and the use of cooling during curing to reduce the temperature gradient. Additionally, this early evaluation of physical characteristics allowed differentiation and measurement of only representative high-quality samples in the final methodology. 

Additional sample photoluminescence emission characterization was completed utilized in the HORIBA FluoroMax-4 spectrofluorometer at Oak Ridge National Laboratory. Measured samples included one bulk photocured two-dye formula, one 3D-printed two-dye formula, and one 3D-printed three-dye formula. All samples were sanded and polished to 12.5 x 10 x 33 mm in order to fit smoothly within a Horiba Fluoromax cuvette holder with the 3D printed samples requiring minimal post-processing. Photoluminescence measurements were taken for each sample over a range of 350 to 600 nm utilizing an excitation wavelength of 300 nm.


\begin{figure}
    \centering
    \includegraphics[width=1\linewidth]{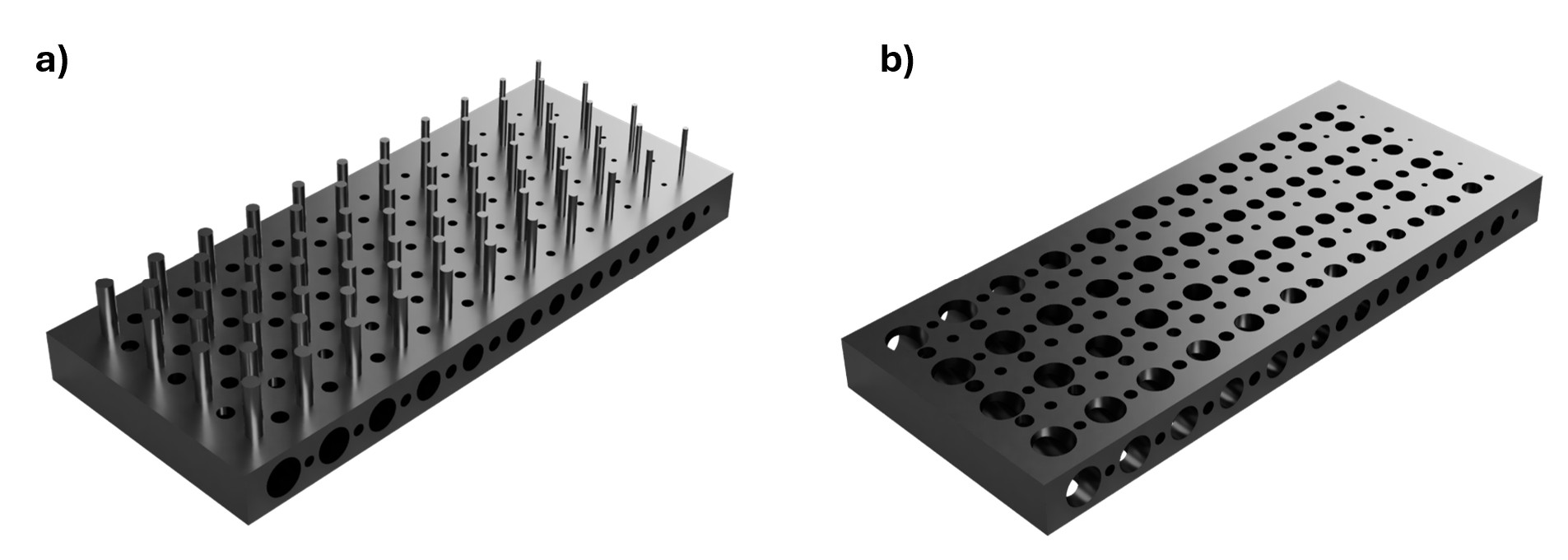}
    \caption{Models for 3D printing resolution test samples featuring a) a model with external pins and internal holes that have diameters ranging from 1.5 mm to 0.5 mm in steps of 0.05 mm, internal horizontal holes with diameters ranging from 3.5 to 1.0 mm in steps of 0.25 and 0.10 mm, and b) a model with internal vertical hole features with diameters from 3.5 to 1.0 mm in steps of 0.25 and 0.10 mm.}
    \label{Res}
\end{figure}


Minimum print resolution was measured using two custom feature-resolution sample designs that were modeled in AutoCAD Fusion 360 and exported to PrusaSlicer for model slicing and conversion into machine-readable g-code. The designs shown in \Cref{Res} contain axial \& lateral holes and axial columns to assess the internal \& external feature resolution limits for the 3D printer and resins. Both are designed at 20 × 60 × 10 mm (L × W × H) to supply sufficient space for internal \& external features from 3.5 mm down to 1 mm in size.  These test objects were printed using both the two- and three-dye formulas and were analyzed through direct measurement of which holes and column pins fully formed. The smallest features resolved, both external and internal, indicate the limits of feature resolution for that resin and printer setting combination. The two formulas were additionally printed in the configuration of F-16 and F-104 aircraft to demonstrate the ability of the scintillator resin to print in applied design structures not easy or possible to replicate with plastic scintillators made by standard subtractive manufacturing.
 
Lastly, printed samples were characterized for their radiation response and compared to the commercially available EJ-200 and EJ-276D plastic scintillators. Printed 1.5" height cylindrical samples were cut to 1" in order to match the geometries of the off-the-shelf scintillators. Samples were printed over-sized and then cut down rather than printed at exactly 1" to quickly remove any excess polymerization in the axial direction along the base of the samples, ensuring identical sample geometry for measurement. Each sample required minimal polishing to 1 micron grit to have a transparent finish. Characterized samples included one bulk photocured two-dye formula scintillator, one 3D printed two-dye formula scintillator, and one 3D printed three-dye formula scintillator. All five samples (3 photopolymerized, 2 off-the-shelf) were coupled with the same Hamamatsu R6231-100 photomultiplier tube (PMT) using Saint-Gobain Crystals BC-630 silicone grease. Each scintillator was prepared for measurement with the wrapping of seven layers of white Mil-T-27730A polytetrafluoroethylene tape around all sides apart from the one that interfaces with the glass window of the coupled PMT. Sample response was recorded using a CAEN DT5730 digitizer with a CAEN DT8033M HV power supply module providing -1,200 V bias voltage. CoMPASS controls data acquisition while GECO performs voltage control. Each scintillator was tested for relative LO with a $^{137}$Cs ($~$ 10 $\mu$Ci) source for 10 minutes at 5 cm. The Compton edge position for each scintillator was determined by finding the half-height of the maximum value of the Compton edge region. Although this light calibration method is known to be biased \cite{Dietze1982}, it is suitable in this case as a means of relative comparison across different scintillators. PSD capability was evaluated through irradiation of the samples with an americium-beryllium source for 15 minutes with light calibration from the $^{137}$Cs measurements. A cadmium foil was placed between the source and detector to reduce the low-energy thermal neutron flux. PSD performance was quantified through a Figure-of-Merit (FoM) value by locating the centroids of the neutron and gamma PSD peaks ($X_n$, $X_{\gamma}$) with their full widths at half maximum ($FWHM_n$, $FWHM_{\gamma}$) and using \cref{FOM}:

\begin{equation}
\label{FOM}
\text{FoM} = \frac{X_\text{n} - X_{\gamma}}{\text{FWHM}_\text{n} + \text{FWHM}_{\gamma}}.
\end{equation}

\section{Results and Discussion}
\label{sec:results}

\subsection{Component Spectral Response}

\begin{table*}[ht!]
\caption{Formula photoactive component peak absorption \& emission}
\label{compspectra}
\centering
\resizebox{\textwidth}{!}{%
\begin{tabular}{@{}cccccccc@{}}
\toprule
 & \multicolumn{2}{c}{PPO} & \multicolumn{2}{c}{Exalite 416} & \multicolumn{2}{c}{Coumarin 450} & TPO \\
 & Absorption & \begin{tabular}[c]{@{}c@{}}Emission \\ ($\lambda_{ex}$ = 303 nm)\end{tabular} & Absorption & \begin{tabular}[c]{@{}c@{}}Emission \\ ($\lambda_{ex}$ = 350 nm)\end{tabular} & Absorption & \begin{tabular}[c]{@{}c@{}}Emission \\ ($\lambda_{ex}$ = 366 nm)\end{tabular} & Absorption \\ \midrule
Literature (nm) & 303, 314 & 336, 354, 370 & 353 & 393, 413 & 366 & 435 & 295, 381 \\
Measured (nm) & 303, 315 & 344, 360, 373 & 350 & 393, 413 & 366 & 436 & 298, 380 \\ \bottomrule
\end{tabular}%
}
\end{table*}

Absorbance and emission spectra for each photoactive component were found to match closely to literature values summarized in \Cref{compspectra} \cite{ Lambdachrome}, \cite{exciton}. To evaluate this comparison, PPO, exalite 416, and coumarin 450 were diluted to 10$^{-5}$ M while TPO was diluted to 10$^{-3}$ M to produce a sufficient spectral response measured in the Agilent Cary 7000 UMS UV-VIS spectrophotometer. For all but one component, the measured peak absorption and emission wavelengths are within 3 nm of literature. The emission peak of the outlier, PPO, is a full 8 nm from expectation. The shift in each of the three PPO peaks indicates impurities or degradation products within the differing samples. Alternatively, slight variations in the excitation source between the measuring devices at this excitation wavelength or solvent purity may contribute to these discrepancies. 

\begin{figure}[ht!]
    \centering
    \includegraphics[width=1.0\linewidth]{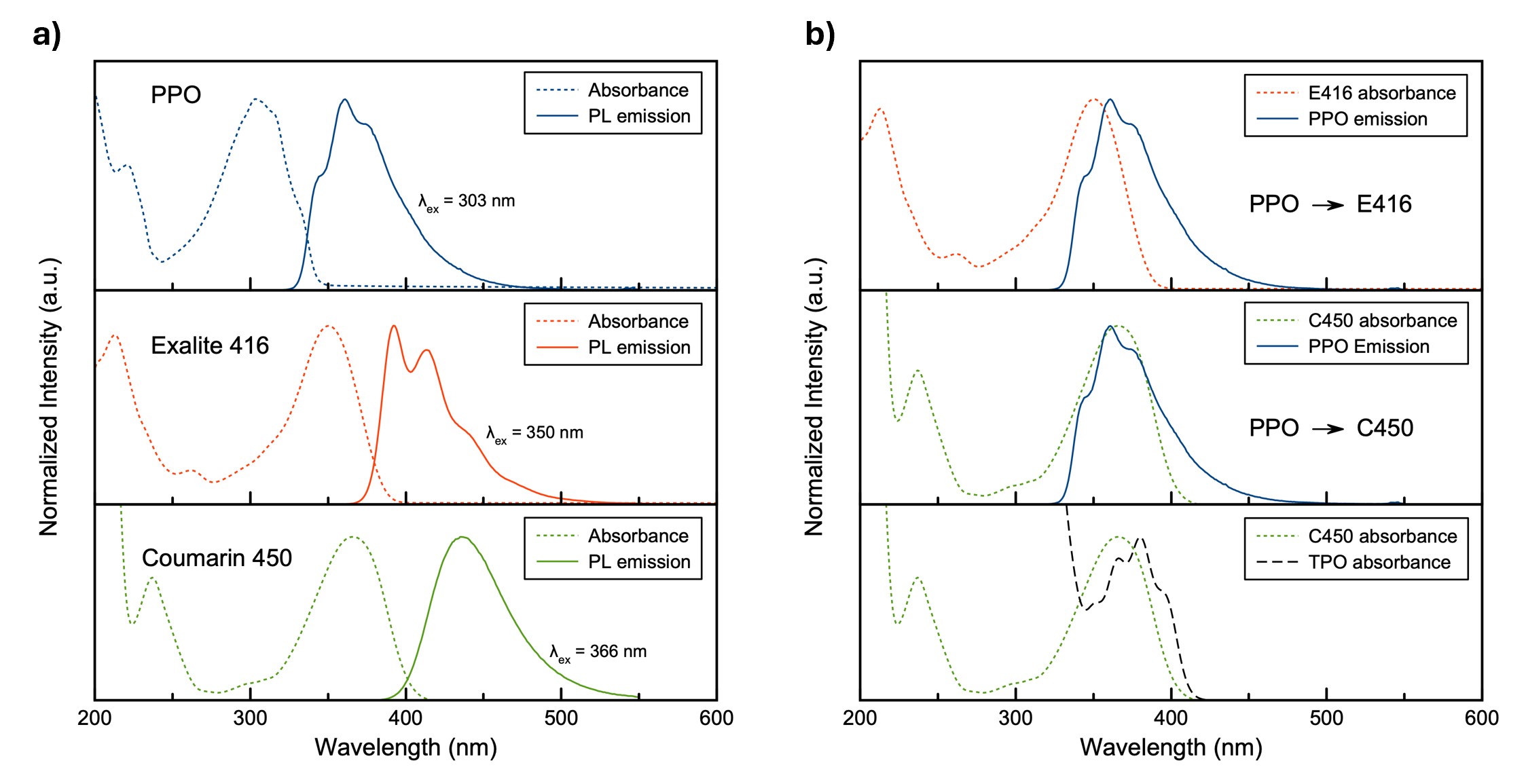}
    \caption{Absorption \& emission spectra for a) each major fluorescent component of the resin formulas and b) the interaction between PPO emission \& exalite 416 absorption, PPO emission \& coumarin 450 absorption, and coumarin 450 absorptions \& TPO absorption.}
    \label{AbsEmiss}
\end{figure}
  
This discrepancy in the PPO emission spectrum does not meaningfully change the relevant interactions between fluors or between the fluors and the photoinitiator. \Cref{AbsEmiss} shows the absorption and emission spectra of each photoactive component as well as overlays between PPO emission and exalite 416 absorbance, PPO emission and coumarin 450 absorbance, and coumarin 450 absorbance and TPO absorbance. Due to the lack of an aromatic base, the mechanism for primary light generation in the scintillator is PPO, which most strongly emits (when optimally excited at 303 nm) between 340 and 370 nm. The peak spectral response of most PMTs lies in the 400–440 nm range, meaning that PPO should not be relied on as the sole fluorescent compound \cite{Hamamatsu}. This indicates the need for additional wave-shifting dyes and is the purpose of the secondary dopant, exalite 416, with absorbance closely overlapping the emission of PPO. Another key interaction is between PPO and coumarin 450. Coumarin 450's emission peak is near 436 nm and much of the emission spectrum is outside the optimal PMT response range. Therefore any light produced from PPO or exalite 416 that gets absorbed by the coumarin compound is likely to worsen the probability of detection in the PMT. The benefit of including coumarin 450 in the formula is due to its interaction with TPO as seen in \Cref{AbsEmiss}b. Coumarin 450 has an absorption spectrum that matches closely to that of TPO, meaning that coumarin 450 competes to absorb the light energy that could go to polymerization instead. Therefore the presence of coumarin 450 limits the penetration of the 405 nm curing light into the liquid resin, which limits the cure depth and enhances the print resolution. A low concentration of the tertiary dye allows for both minimization of the harmful competition with exalite 416 while also enabling cure depth control.

\subsection{3D-Printed Scintillator Fabrication}

\begin{figure}
    \centering
    \includegraphics[width=1\linewidth]{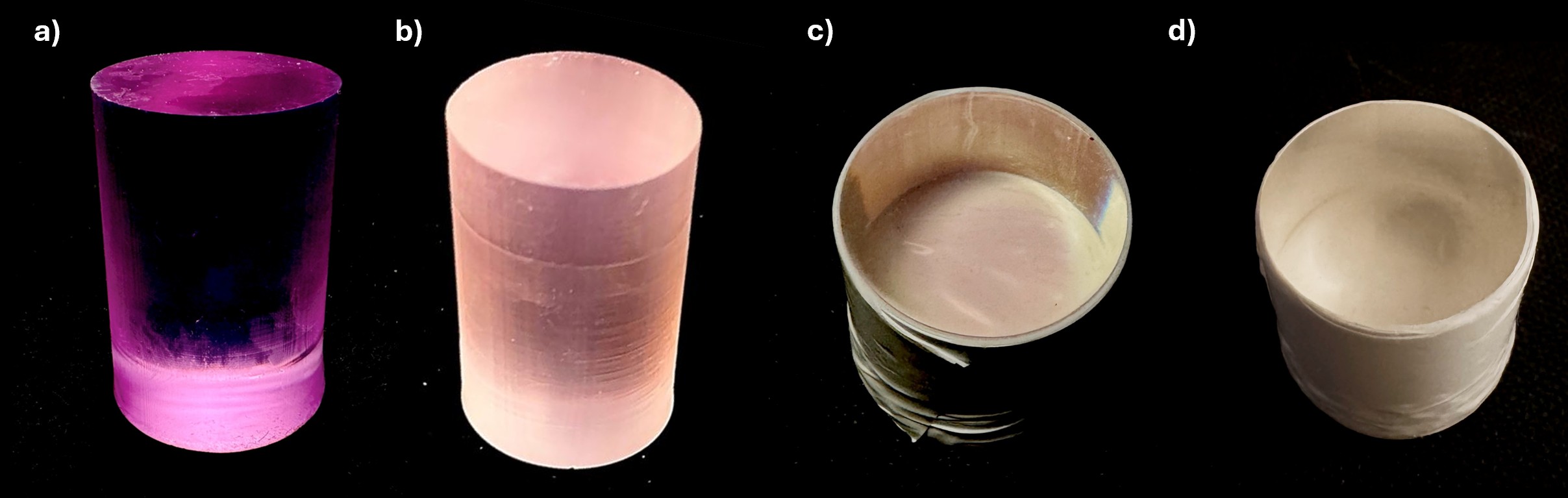}
    \caption{Decay of purple discoloration over 14 days, shown by a) a 3D printed scintillator immediately after printing, b) the same geometry 3 days after printing, c) the now cut and polished geometry 7 days after printing, and d) the same scintillator 14 days after printing with no purple discoloration.}
    \label{Purpling}
\end{figure}

A significant purple discoloration was observed in all scintillators upon exposure to curing light from the Prusa CW1S or Prusa SL1S 3D printer. Previous work \cite{Frandsen2023} found that this discoloration increased with the amount of PPO in both intensity and duration, ranging from minutes to weeks at room temperature. In this study, all scintillators contained an equivalent amount of PPO (20 wt.\%), and while the method of initial curing differed between samples, all samples were exposed to the Dymax FX-1250 for burn-up of TPO. This post-curing method resulted in all samples requiring 14 days at room temperature for purple discoloration to dissipate, as shown in \Cref{Purpling}.

The purpling is theorized to result from the interaction of PPO and TPO in the presence of UV and near UV light, and therefore it is only visible where significant curing has occurred. Previous work has additionally found that the dissipation of purple color can be accelerated using an oven and elevated temperatures \cite{Frandsen2023}. This study did not explore these options to mitigate the risk of additional discoloration from heat. 

\begin{figure}[ht!]
    \centering
    \includegraphics[width=1\linewidth]{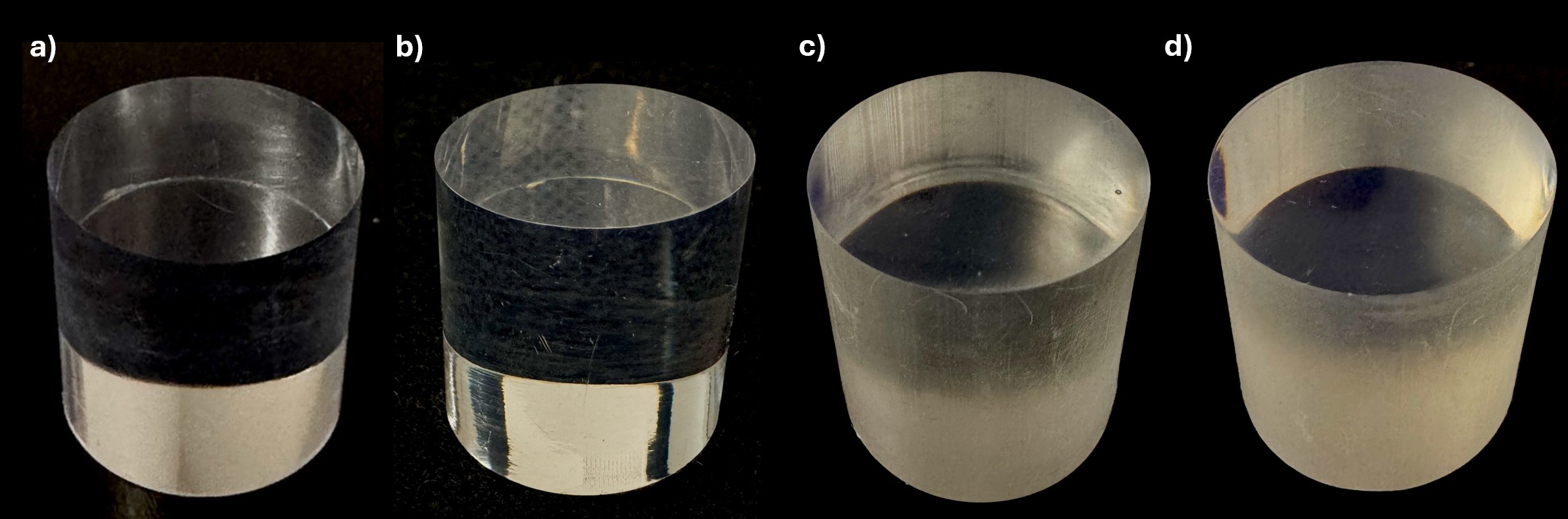}
    \caption{Comparison of final bulk photocured and 3D printed scintillators: a) a commercially thermal cured EJ-200 for comparison, b) the two-dye formula after bulk curing and polishing, c) the same two-dye formula after 3D printing and polishing, and d) the three-dye formula after 3D printing and polishing.}
    \label{Yellow}
\end{figure}

\Cref{Yellow} shows four samples utilized in final radiation detection and spectral analysis. While the two-dye sample (sample b) is equally transparent to the EJ-200 (sample a), it also has a very faint yellow color. Samples c and d represent the 3D printed versions of the two-dye and three-dye formulas, respectively. A slightly more intense yellow coloration is present in these samples. The yellow color results from several sources. The first is chemical composition. The base oligomer that makes up the bulk of the formulas has a Pt-Co (APHA) color of 15 \cite{Bomar}, which indicates that while not significant, yellow coloration is inherent to the bulk of the mixture. Also, the photoinitiator used within the samples (TPO) at 0.1 wt.\% is a pale yellow powder, further adding yellow coloration to the samples. 

\begin{figure}[ht!]
    \centering
    \includegraphics[width=.6\linewidth]{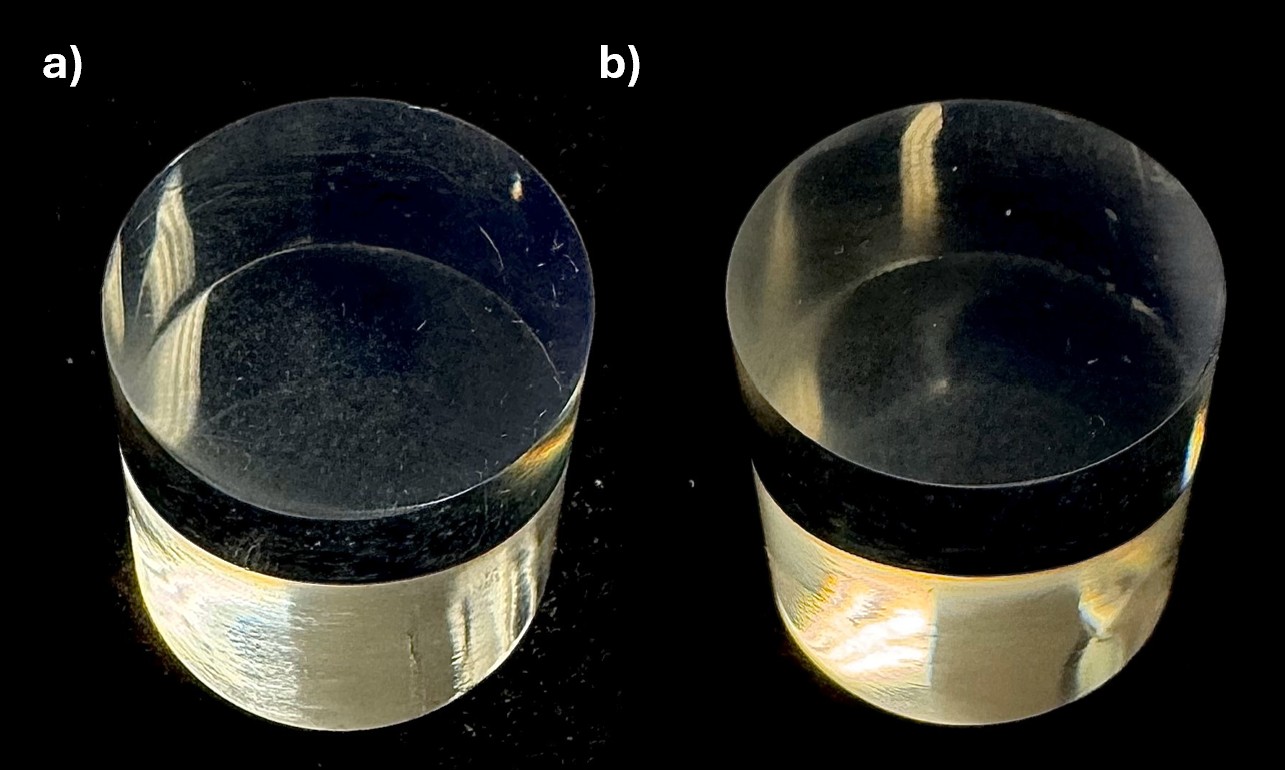}
    \caption{Scintillators after bulk photocuring while a) keeping sample temperature low and b) allowing the temperature to exceed 60 $^{\circ}$C.}
    \label{Heat}
\end{figure}

 
Heat and overexposure to UV light during post-curing also caused samples to yellow. \Cref{Heat}a and b show samples from an identical batch and formula cured using bulk photocuring. The sample in \Cref{Heat}a was cooled during the curing process using a bath of ice water and the sample in \Cref{Heat}b had no cooling, allowing it to reach temperatures near or above 60 $^\circ C$. Yellow discoloration is present in the uncooled sample, indicating that additional heat acts to thermally oxidize the polymers during or after polymerization or otherwise damage or form chromophores in the samples. This heating is especially problematic when 3D printing samples, as shown by the increased discoloration seen in \Cref{Yellow}c and d, because of the significant amount of heat produced within the sample layers and the vat containing the liquid resin during the repeated exposure and polymerization from layering. This rapid heating leads to the vat of liquid resin reaching over 60 $^\circ C$ and causing printed geometries to begin to yellow. Modified printing settings to increase rest times between layers, increase fan speeds, and pause the print once the measured temperature reaches 50 $^\circ C$ were implemented and found to reduce yellowing in printed samples. 


\begin{figure}[ht!]
    \centering
    \includegraphics[width=0.55\linewidth]{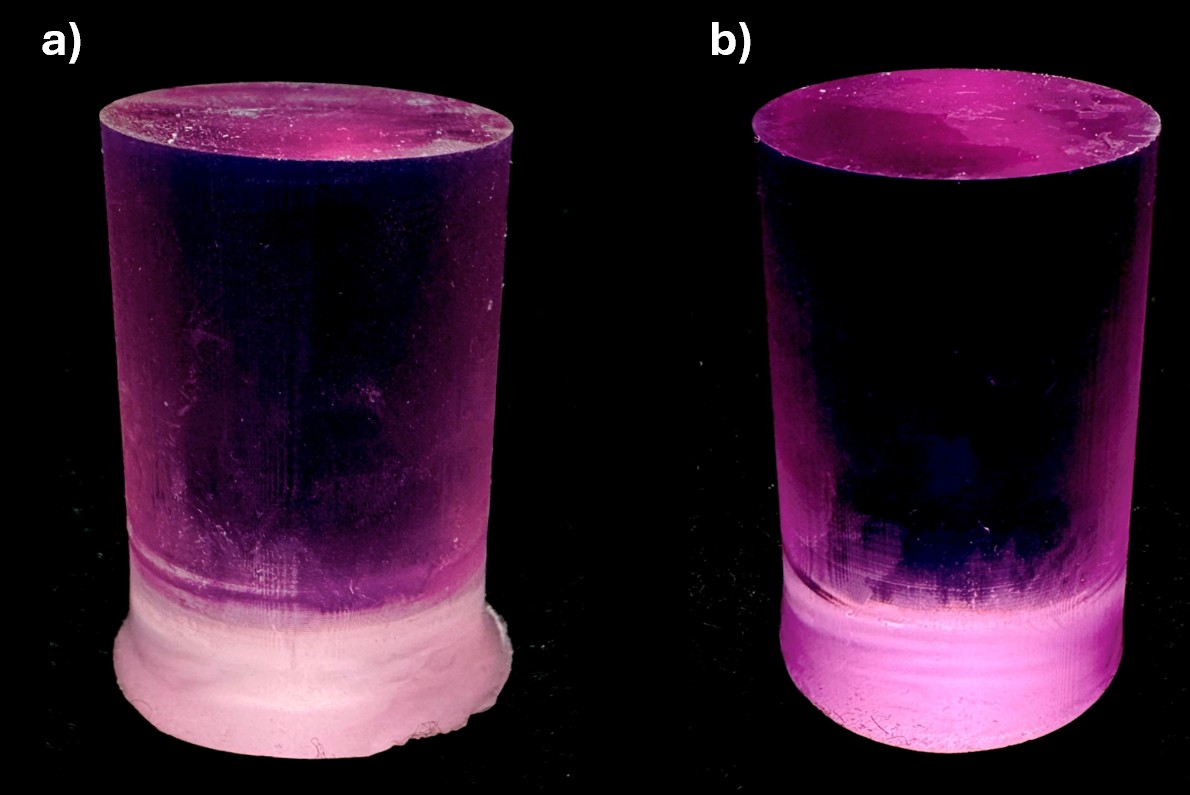}
    \caption{Comparison of increased excess polymerization of a) two-dye formula printed 1.5" cylindrical geometry scintillator at 45 s/layer to b) three-dye (coumarin 450) formula printed 1.5" cylindrical geometry scintillator using identical settings.}
    \label{3DPrintedScint}
\end{figure}

The use of the tertiary dye coumarin 450 substantially mitigated unintended polymerization in monolithic samples, particularly along the base of the geometry cured to the build plate. Scintillators printed with the two-dye formula, such as \Cref{3DPrintedScint}a, show a significant degree of excess polymerization while scintillators printed with the three-dye formula, such as \Cref{3DPrintedScint}b, show none. Both of the scintillators in \Cref{3DPrintedScint} were printed at an identical 45 s/layer and post-cured for a similar 5 minutes. While for simple 1" cylinders this section of the geometry can be cut off, the presence of extraneous polymerization indicates a lack of resolution of printed geometry at a macroscopic scale. The tertiary dye, on the other hand, accomplishes its purpose in limiting the depth for which the curing 405 nm light can reach into the resin and cause polymerization beyond the designed features. 

\subsection{Plastic Scintillator Spectral Response}

\begin{figure}[hb!]
    \centering
    \includegraphics[width=0.5\linewidth]{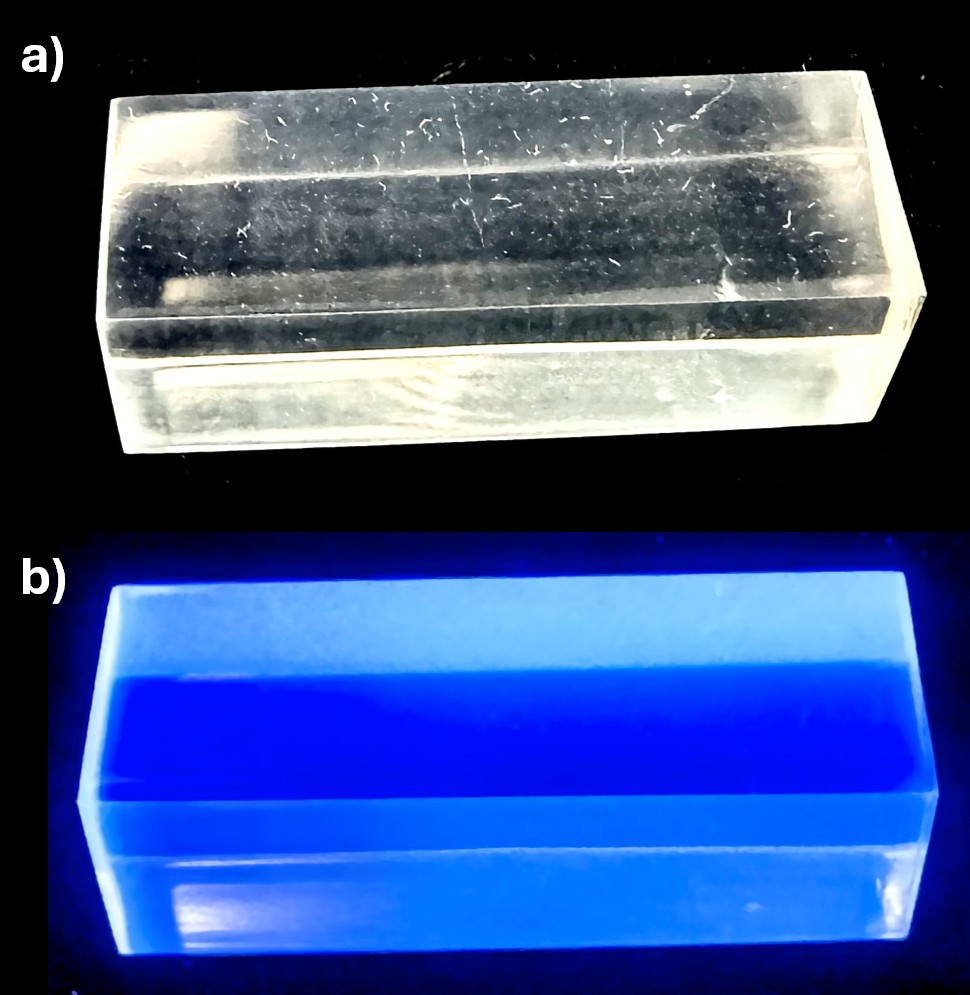}
    \caption{Scintillator formula printed in 12.5 x 10 x 33 mm cuvette geometry under a) normal light and b) 405 nm excitation light.}
    \label{CuvetteCure}
\end{figure}

\begin{figure}[hb!]
    \centering
    \includegraphics[width=0.8\linewidth]{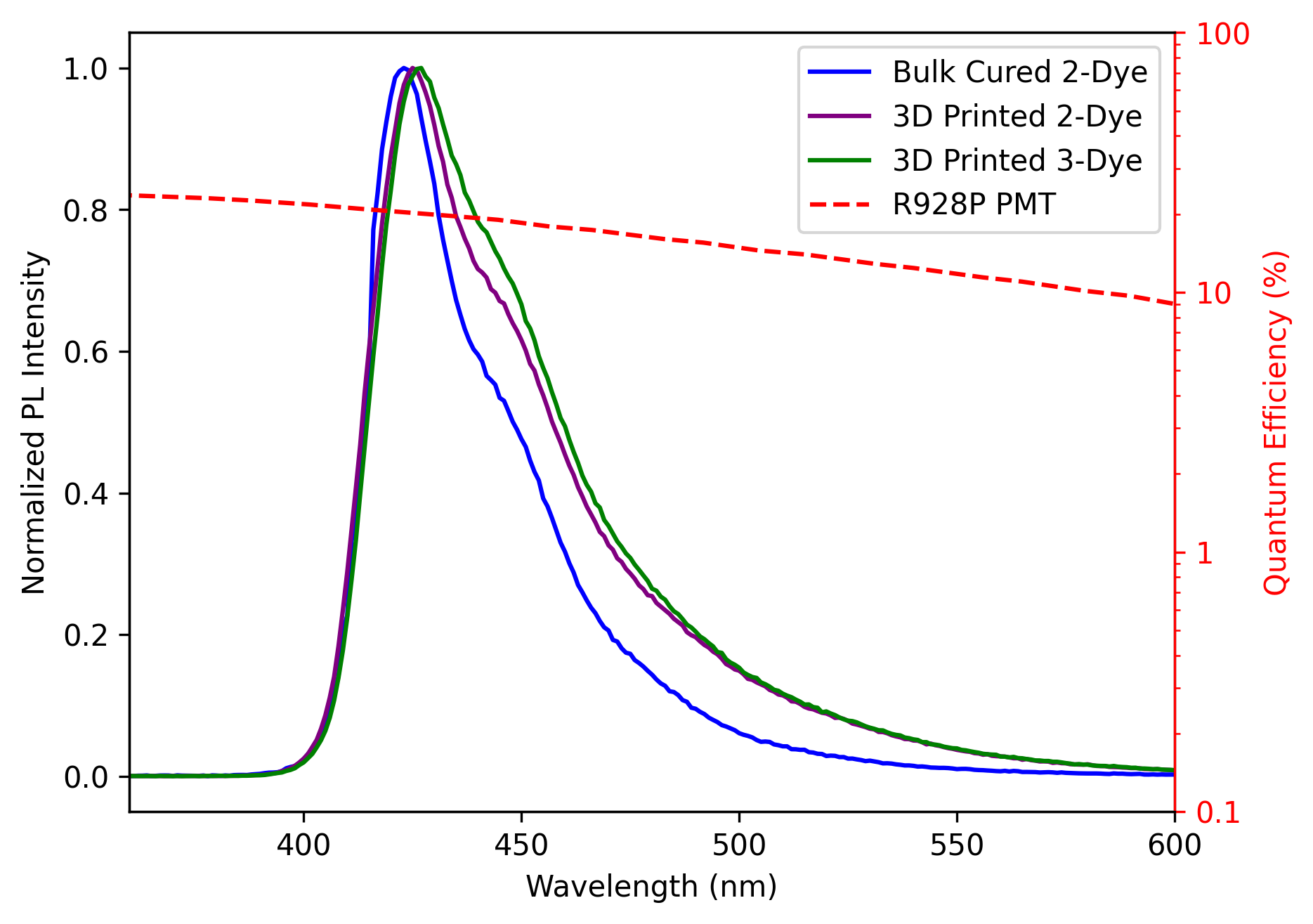}
    \caption{Photoluminescence spectral response of three formulas to 300 nm excitation light and normalized to peak emission and quantum efficiency of R928P PMT in red \cite{hamamatsu_R928}.}
    \label{Photolum}
\end{figure}

\Cref{Photolum} shows the photoluminescence spectra of the three formulae manufactured in 12.5 x 10 x 33 mm cuvette geometry shown in \Cref{CuvetteCure} closely match and all most intensely emit between 420 and 430 nm. These curves also closely match the expected emission of exalite 416 from \Cref{AbsEmiss}, indicating the pathway of light transfer from primary fluor to secondary dye is functional and remitting it at the longer wavelength as required for high quantum efficiency with PMTs. This utility is further supported by the limited intensity below 400 nm. The dashed red line plotted along the second axis of \Cref{Photolum} represents the quantum efficiency of the R928P PMT utilized in HORIBA FloroMax 4 spectrofluorometer used for photoluminescence measurement. This curve's overall flat quantum efficiency over the wavelength range of interest indicates that most of the light produced by PPO is reabsorbed and the reduced intensity is not simply a result of low PMT absorption efficiency at this wavelength range. This PMT differs in quantum efficiency compared to those typically used for scintillator radiation detectors. Small differences were observed in the photoluminescence spectra of the cured plastics depending on the inclusion of the tertiary dye and to a lesser extent on the curing method. Of note is that the bulk-cured two-dye formula sample has the narrowest spectrum, indicating that the 3D printing of the samples has caused a broadening of the emission pattern of the samples. As stated previously, the 3D printed samples are slightly more yellow, indicating more absorption of the emitted scintillation light. The 3D printed three-dye formula is also rightward shifted and broader than the two-dye formula samples, indicating that despite its low concentration, the tertiary dye absorbs enough of the other compound emissions that a visible response in the overall emission of completed samples is observed. However, at least for 300 nm excitation, this effect is small, and peak emission is still within the region of high collection quantum efficiency for PMTs due to the low concentration of the tertiary dye.


\subsection{Feature Size Limitations}

The optimal resolution control formula was found to use a concentration of 0.04 wt.\% coumarin 450 and a per-layer curing time of 30 s/layer. To determine these optimal parameters, layer curing times were evaluated at 20 s/layer, 25 s/layer, and 30 s/layer. This range of curing times were chosen through analysis of the resin during bulk polymerization, where the liquid resin solidified between 15 and 30 seconds. Additionally, layer curing times below 20 seconds caused the print to fail while layer curing times above 30 seconds caused unnecessary excess polymerization. Dye concentration ranged from 0 (no tertiary dye) to 0.1 wt.\%. A concentration of 0.1 wt.\% was chosen as the highest concentration because samples above this concentration did not fully cure, causing the following layers to improperly bond. All feature resolution test objects were printed with a layer height of 0.1 mm, resulting in production times for the external feature resolution device being 37, 45, and 54 minutes for the 20, 25, and 30 s/layer prints, respectively. As the internal feature resolution test objects did not have additional vertical pins, these printed faster at 18, 23, and 27 minutes for the three-layer curing times, respectively. 

\begin{figure}[ht!]
    \centering
    \includegraphics[width=1.0\linewidth]{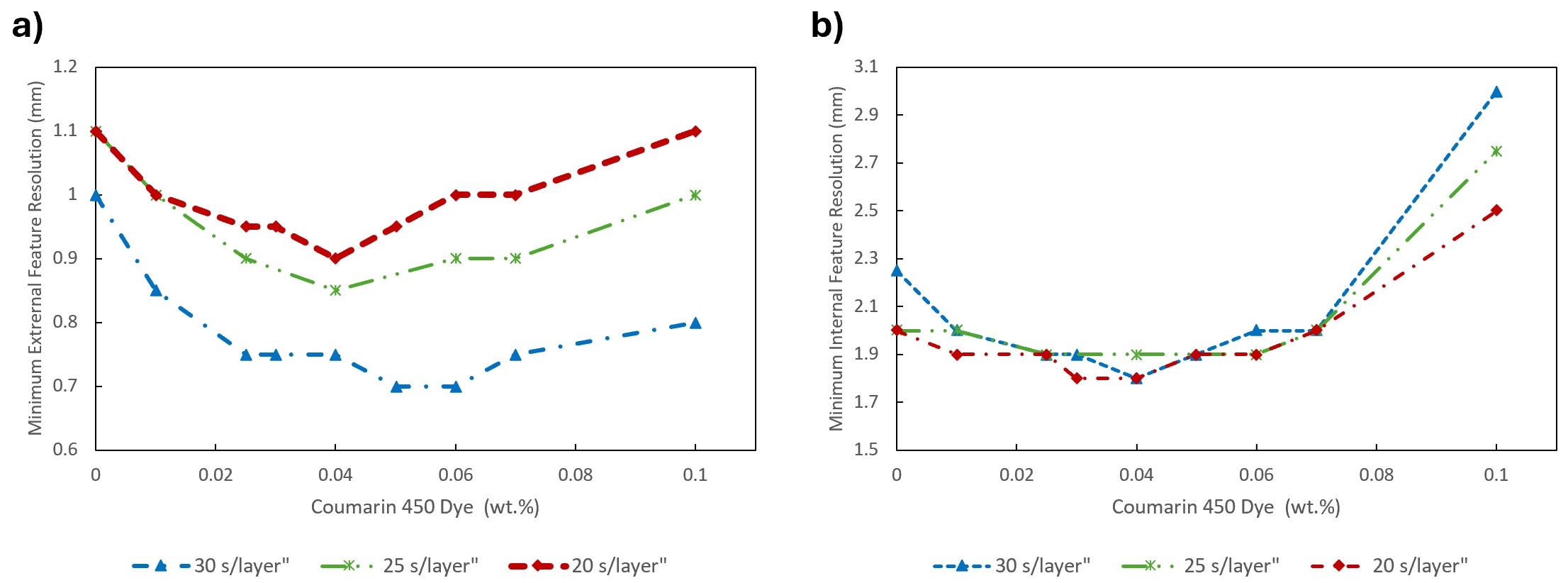}
    \caption{Minimum achievable a) external and b) internal resolution at variable concentration of coumarin 450 cure depth limiting dye and layer curing time for a set 0.1 mm curing layer thickness. 0 \% represents the two-dye formula using no tertiary dye.}
    \label{ResPlot}
\end{figure}
 
\Cref{ResPlot}a shows the minimum external feature resolution achievable for the three curing times with all three curves trend downward with increasing concentrations of cure depth dye up to a certain point. After this inflection point, the lines trend upward. This change in trend suggests that increasing concentrations of tertiary dye initially enhance feature resolution by limiting the amount of stray curing light that can scatter within the resin or reflect off geometric surfaces, reducing unintended partial curing around the desired features. At the same time, the curing light is concentrated within a narrower region, leading to improved curing precision and finer resolution. At the inflection point, the dye absorbs enough of the polymerization-initiating light so as to significantly reduce the overall energy available for curing. For the set 0.1 mm layer thickness, this lower energy means thinner layer thickness and incomplete curing such that the structural integrity of the external features is weakened to the point that they cannot support further layering, preventing smaller features from resolving. This inflection point for a 0.1 mm layer thickness occurs at 0.04 wt.\% coumarin 450 for the 20 and 25 s/layer curing trials and between 0.05 and 0.06 wt.\% coumarin 450 for the 30 s/layer trials. The curves shift downward with increased time, meaning smaller axial pins are better resolved for larger per-layer curing times. For the short per-layer curing times, insufficient curing light initiates polymerization within the resin to fully solidify the layer. This insufficient light leads to subsequent layers either weakly bonding to the former layer, creating interlayer defects, or simply not bonding at all, causing the feature not to appear. Comparing the curves, the minimum external feature resolution possible with the current set of printer settings and resin tested is 0.7 mm at a per-layer curing time of 30 s/layer and a dye concentration of 0.05 or 0.06 wt.\% coumarin 450. 

\Cref{ResPlot}b shows a similar trend for internal feature resolution where all three layer curing time curves trend downward, indicating improvements in the minimum internal feature resolved followed by an inflection point beyond which additional tertiary dye drastically reduces the smallest feature resolvable. For internal features, this inflection point for all three layer curing times appears to be 0.04 wt.\% coumarin 450. Before this point, adding the dye improves the resolution as the absorption of the curing light prevents sufficient light from reaching beyond the set boundaries within and beyond the layer, preventing vertical and horizontal over-curing. However, beyond the inflection point, additional tertiary dye absorbs too much of the light, so insufficient curing energy can be delivered through the full thickness of the layer. This results in the following layers not fully adhering, meaning that each semi-cured layer falls into previous layers and prevents the internal feature from being fully resolved. For the internal features, minimal difference in achieved minimum resolution between the three per-layer curing times can be seen with the shortest curing time only resolving at most one additional feature. The observed minimum achievable resolution for all three curing times is 1.8 mm at 0.04 wt.\% coumarin 450. If this is taken into account with the external feature resolution plot in \Cref{ResPlot}a, the optimal resin given the printer and resin parameters tested is a layer curing time of 30 s/layer and a 0.04 wt.\% concentration of tertiary cure depth dye, coumarin 450. This concentration of tertiary dye produces the best internal feature resolution with only a small reduction in external unsupported feature resolution. 

\begin{figure}[ht!]
    \centering
    \includegraphics[width=0.85\linewidth]{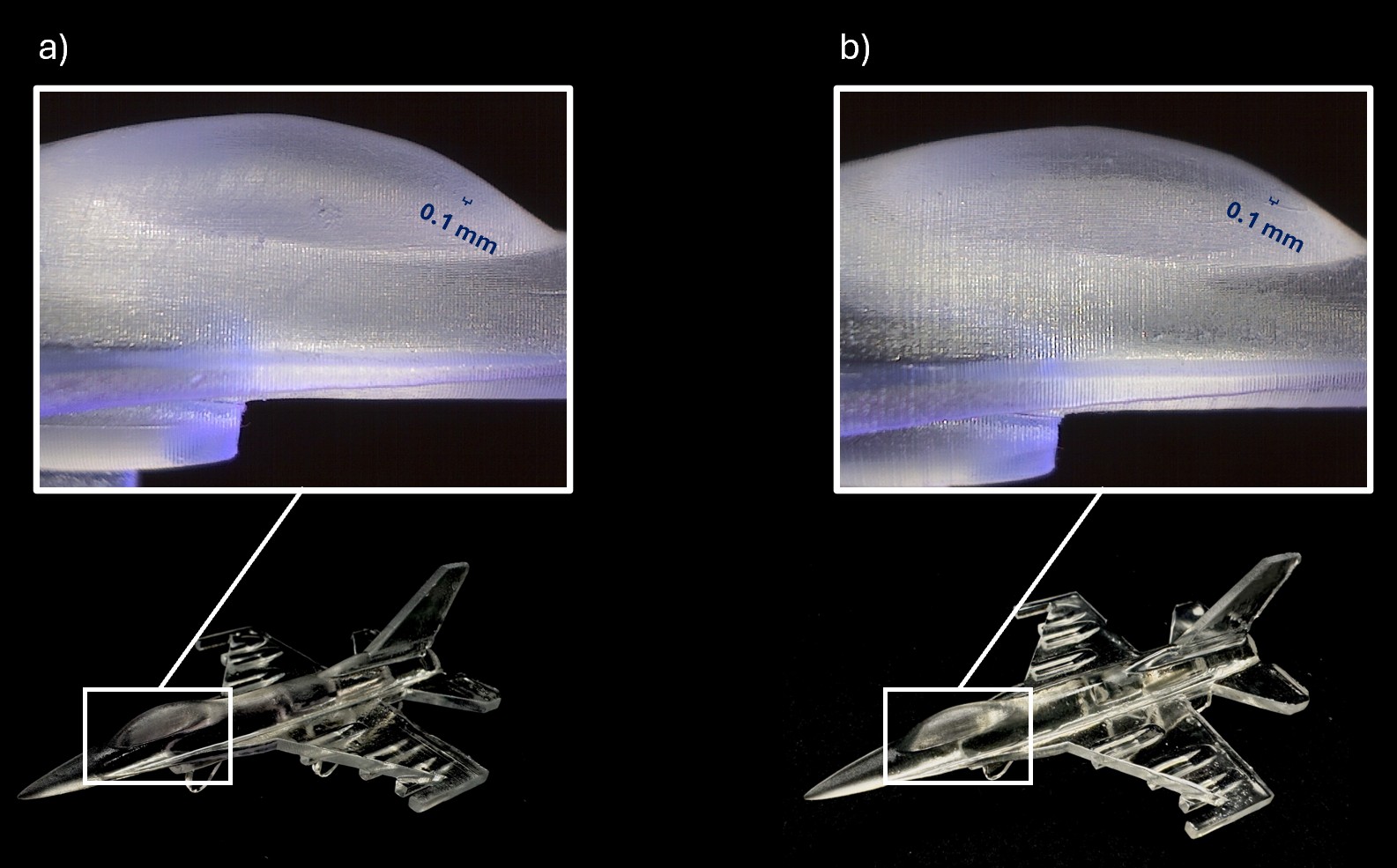}
    \caption{Complex feature geometry F-16 printed using a) two-dye formula showing feature resolution around the 0.1 mm level with a small amount of inter-layer blurring observed in zoomed-in view of canopy and b) three-dye scintillator formula showing precise small feature resolution below the 0.1 mm level with minimal inter-layer blurring observed in zoomed in view of canopy.}
    \label{F16}
\end{figure}

\subsection{Complex Feature Geometry}



\begin{figure}[ht!]
    \centering
    \includegraphics[width=.85\linewidth]{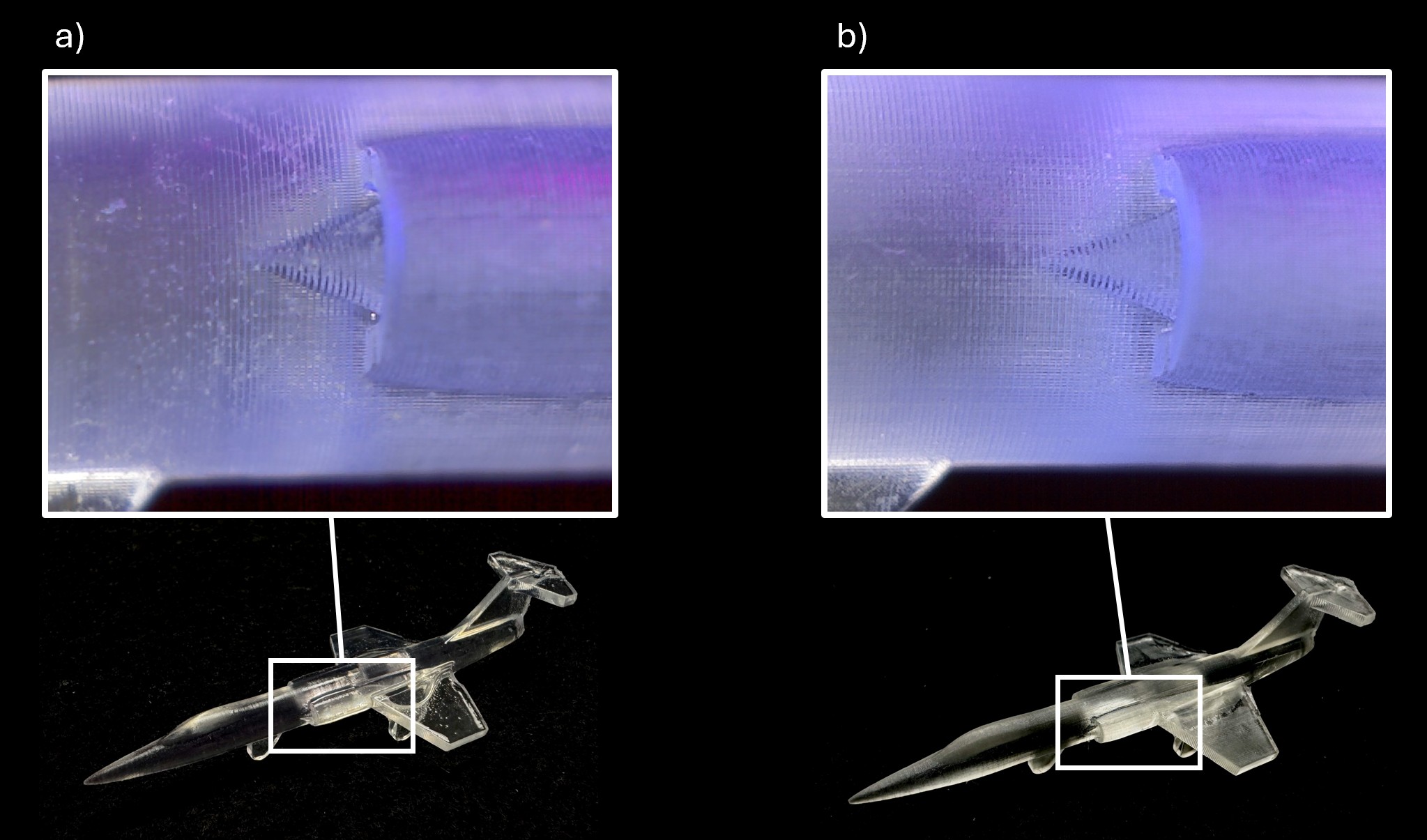}
    \caption{Complex F-104 geometry printed using a) two-dye formula showing feature resolution around the 0.1 mm level with zoomed-in view showing turbine rotor with small amount of inter-layer blurring and b) three-dye scintillator formula showing precise small feature resolution below the 0.1 mm level with zoomed in view showing minimal to no inter-layer blurring in the turbine rotor.}
    \label{F104}
\end{figure}

Beyond simple 1" cylinders and feature resolution test objects, several complex geometries were fabricated. \Cref{F16} and \Cref{F104} show prints of custom-designed F-16 and F-104 geometries using both two-dye and three-dye formulas. Each was printed using 100 $\mu$m layers and 30 s/layer curing times with the SL1S 3D printer at the designed 2.2 mW/cm$^2$ 405 nm intensity. These settings resulted in a manufacturing time of 5 hours and 31 minutes for the designs to print. Each additionally used breakaway support structures beneath the wings and tail architecture that allowed the printing of high-angle and fully horizontal layers that would otherwise be impossible to manufacture in vat-based stereolithography. Typical resin 3D printing requires initial ``burn-in'' layers, consisting of the first 5–10 layers, to have significantly longer per-layer curing times in the range of 5 – 10 times all subsequent layers to ensure a strong adhesion to the build plate. This resin, however, was found to adhere to the build plate in as little as 20 seconds and so all layer curing times were identical.

Geometries printed using the two-dye formulation are shown in \Cref{F16}a and \Cref{F104}a, while those printed with the three-dye formulation are shown in \Cref{F16}b and \Cref{F104}b. The three-dye prints exhibit improved small-feature resolution and reduced excess polymerization and inter-layer blurring. All four geometries demonstrate the ability to resolve sub-millimeter features—such as the equipment pods on the F-16 and the jet inlets on the F-104. However, in the geometries printed with the two-dye formulation, excess polymerization is evident. 

\begin{figure}[ht!]
    \centering
    \includegraphics[width=.9\linewidth]{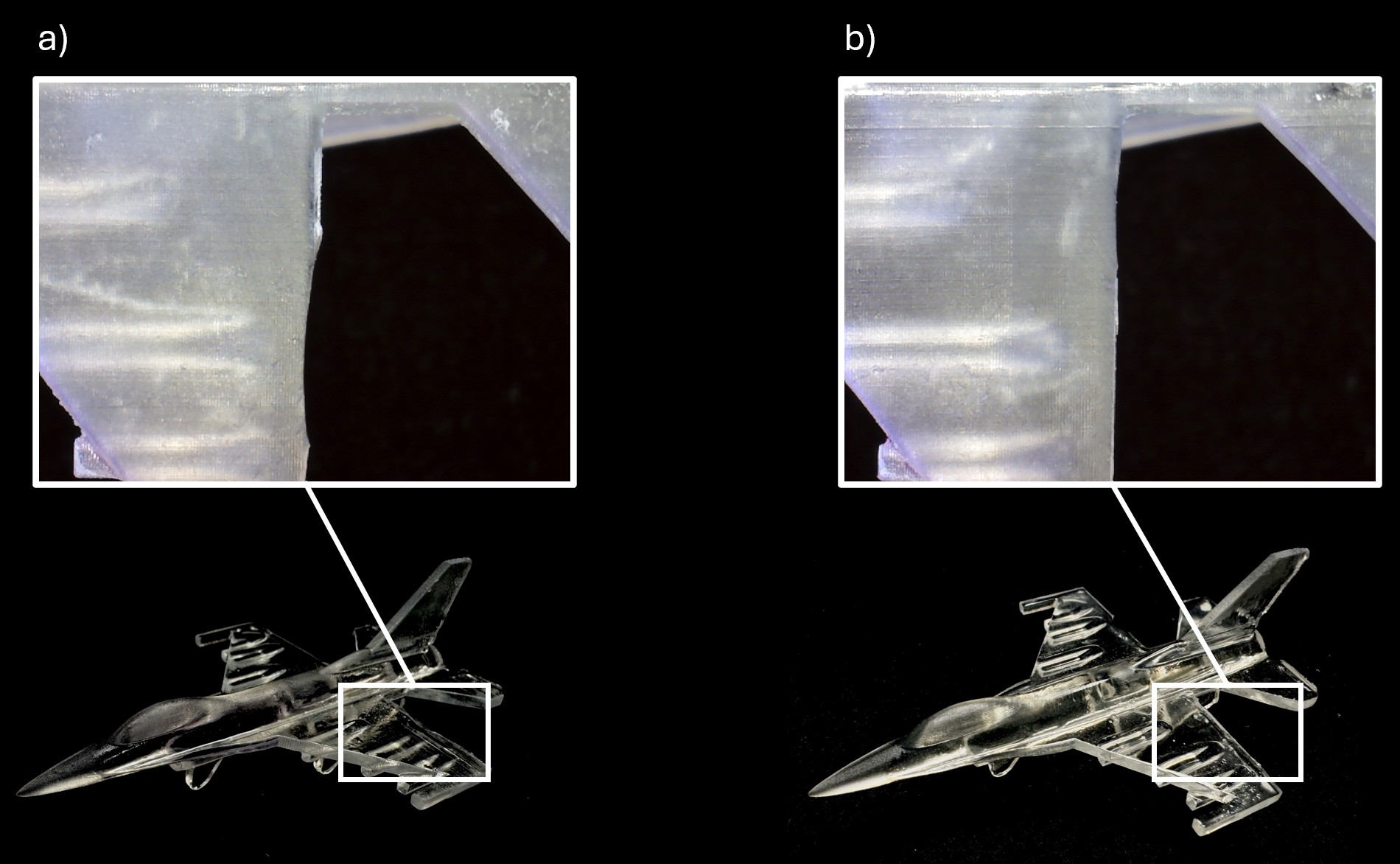}
    \caption{Complex feature geometry F-16 printed using a) two-dye formula showing excess polymerization in curved trailing edge of wings and b) three-dye scintillator formula showing straight trailing wing edge with minimal to no excess polymerization.}
    \label{F16_wing}
\end{figure}

In \Cref{F16}a, \Cref{F104}a, and \Cref{F16_wing}a the trailing edges or ailerons of the wings extend beyond their intended flat profile, a sign of unintended curing. This over-curing also causes layers to blur together, as seen in the zoomed-in canopy and wing views, suggesting prints made using the two-dye formula reach a maximum resolution at or larger than the 0.1 mm print layer height. This layer blurring is consistent with the behavior observed in the resolution test objects, where light penetrates too deeply into the resin, polymerizing regions beyond the intended exposure region. In contrast, excess polymerization and layer blurring are significantly diminished in \Cref{F16}b, \Cref{F104}b, and \Cref{F16_wing}b enabling clear resolution of the curvature of the canopy layers in \Cref{F16}b indicating a maximum feature resolution below 0.1 mm.

Furthermore, by reducing excess polymerization and layer blurring using the tertiary dye, even smaller features can fully resolve in these geometries. This includes the drag chute of the F-16 and the turbine rotor of the F-104 shown in \Cref{F104}b. Additional prints using layer heights down to 25 $\mu$m and curing times down to 15 seconds were attempted but found to produce samples of lower resolution quality and physical stability.

\subsection{Scintillator Performance}

Although the 3D printing process was found to affect scintillator light output and PSD capability, the use of tertiary dye had a negligible impact. All samples in polished 1" cylinder geometry were evaluated for their overall scintillation properties, including relative LO and PSD capability. 

\begin{figure}[ht!]
    \centering
    \includegraphics[width=0.8\linewidth]{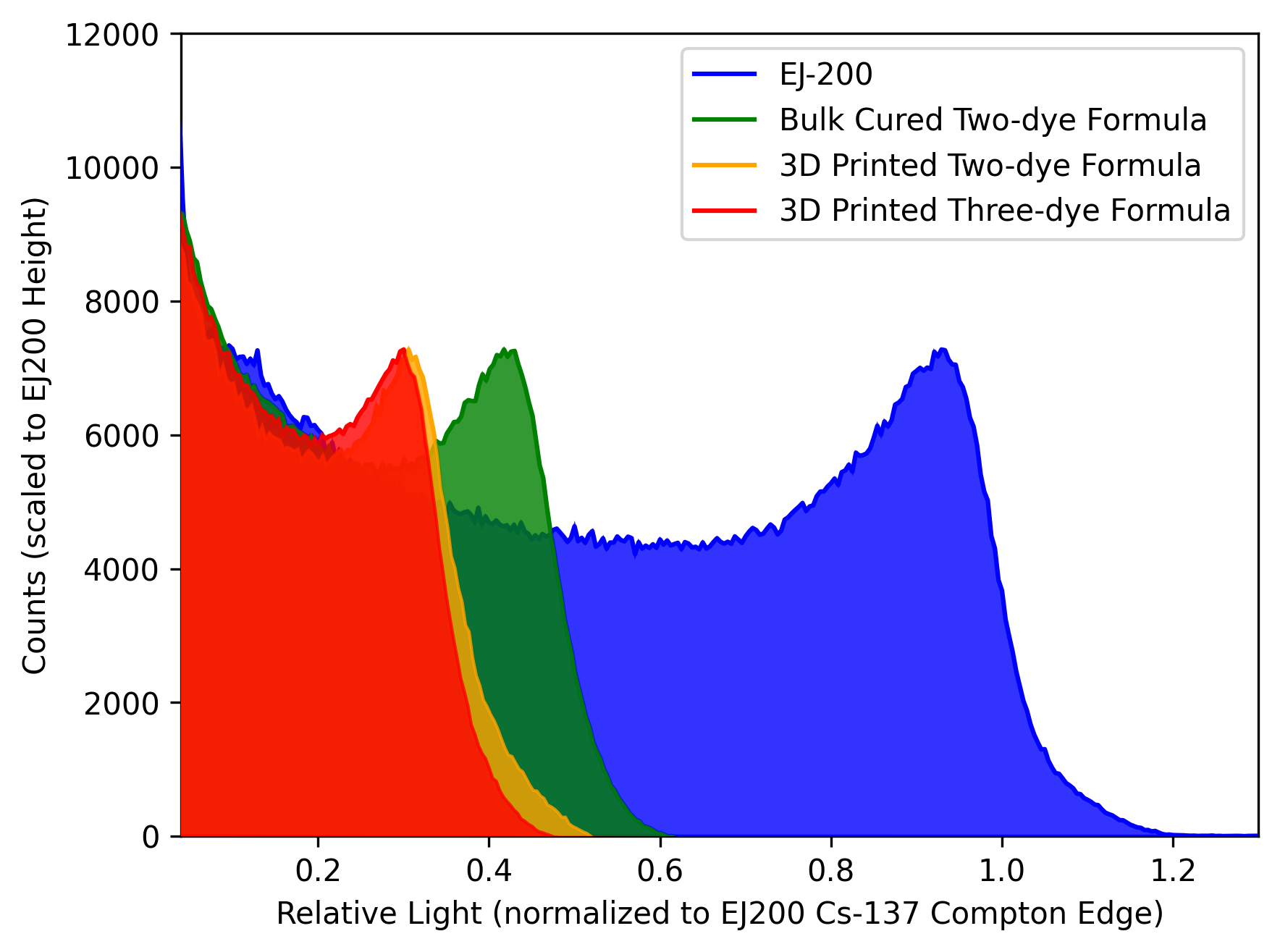}
    \caption{Relative light histogram response comparison using $^{137}$Cs at 5 cm distance for two- and three-dye formula resin compared against EJ-200 in 1" cylindrical geometry.}
    \label{EHist}
\end{figure}

The light outputs of the photocured formulas were compared to an identically sized commercially available scintillator (EJ-200) using a $^{137}$Cs source. \Cref{EHist} summarizes these results with the Compton edges calibrated relative to the EJ-200 edge. The bulk photocured sample response's Compton edge in green is 48.5\% that of EJ-200. This suppressed light yield results from the formula base monomer/oligomer being an acrylate and non-aromatic, resulting in essentially all the LO coming from the primary fluor PPO, which only makes up 20\% of the formula. In comparison, the industry standard EJ-200 base monomer is aromatic, meaning nearly the entirety of the formula is active in the scintillation process. While the light output is reduced, the 50\% performance is high when compared to the 20\% scintillating fraction: as originally shown by Frandsen, et al., a substantial percentage of the performance for a scintillator can come from a small percentage of radiative fluor ingredients \cite{Frandsen2023}. 


The 3D printed two-dye and three-dye formulas represented in orange and red have relative LOs of 36.4 and 35.2 \% of EJ-200, respectively. Because the 3D printed two-dye sample was made with an identical formula and batch as the bulk photocured sample, the lower light output of the 3D printed sample indicates that optical effects are disrupting the light collection efficiency. Light is likely reflecting within individual layers of the printed sample. Alternatively, the slightly yellower color from the heat of 3D-printing the samples may be causing absorption, resulting in a suppressed light output. Some of the observed light suppression may also result from the 3D printing process introducing oxygen, which is a well known cause of quenching \cite{Birks}. While the resins are degassed under vacuum, the 3D printer and process are done in ambient atmospheric conditions. Oxygen can be introduced both during the transfer of resin into the 3D printer vat and during every layer where the vat is tilted. The 3D-printed three-dye formula sample is suppressed slightly more than the two-dye formula. This slight difference indicates that the tertiary dye absorbs a small amount of the emission light as theorized from the absorption and emission plots in \Cref{AbsEmiss}. However, the effect is small due to the very low concentration of the coumarin 450 dye.  

\begin{figure}[hb!]
    \centering
    \includegraphics[width=1\linewidth]{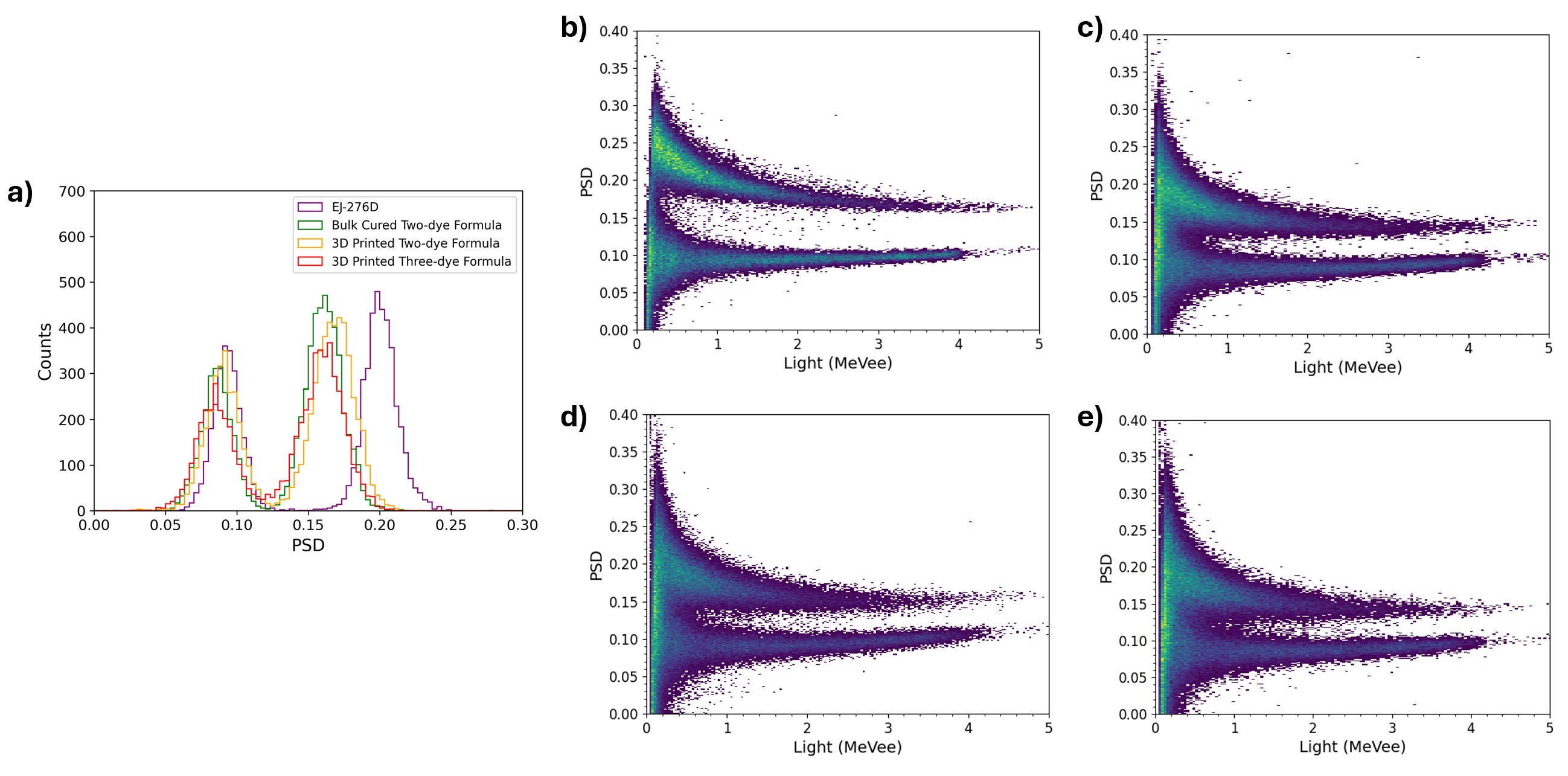}
    \caption{Comparison of a) 1D and b)-e) 2D PSD spectra using an americium-beryllium source for three formulas as well as EJ-276D. 2D spectra are shown for b) EJ-276D, c) two-dye formula when bulk photocured, d) two-dye formula when 3D printed, and e) three-dye formula when 3D printed.}
    \label{PSD}
\end{figure}

All photocured and 3D printed samples showed PSD capability based on the separation of neutron and gamma events when exposed to an americium-beryllium source as demonstrated in \Cref{PSD} and quantified using FoM. While there is no definitive sufficient FoM value, literature shows that a FoM value above 1.3 is adequate for most discrimination applications \cite{Lint}. The FoM generally depends strongly on the chosen light cut. A low light region between 0.2–0.4 MeVee may yield an insufficient FoM, while a higher light region above 2 MeVee can produce a substantially greater FoM. This variation highlights the importance of clearly defining the light cut range for literature comparisons. Moreover, the use of higher energy neutron sources, such as DT generators, for certain applications can reduce the importance of precise low-energy separation. For these measurements, a lower middle portion of the plot between 0.9 and 1.1 MeVee was chosen for the light cut range. A new Eljen EJ-276D (the industry-standard PSD plastic) was compared against the bulk cured and 3D printed samples \cite{Eljen}. This EJ-276D scintillator was maintained under nitrogen atmosphere until immediately before testing took place.

The FoM values for the measured samples over the 0.9 – 1.1 MeVee cut were 2.22, 1.35, 1.29, and 1.20 for the EJ-276D, bulk photocured two-dye, 3D printed two-dye, and 3D printed three-dye samples respectively. \Cref{PSD}a shows the comparison of the 1D PSD projections using the aforementioned light cut. While the EJ-276D has a larger FoM value, the separation between event types for all samples is discernable and each FoM value is above or just below the 1.3 threshold. The reduction in separation performance for the bulk photocured two-dye formula results from reasoning analogous to the case of reduced light yield. A PSD-capable dye allows for a large FoM even at low light in thermally cured samples where the base is aromatic and inherent LO is high. The separation between event types is reduced with acrylate-based formulas such as the ones used in this work, as the base is non-aromatic and relative LO is lower. The EJ-276D was stored under nitrogen prior to use and thus had minimal oxygen or light exposure, while the photocured formula (which was degassed under vacuum) was cured in ambient air, allowing air to diffuse back into the solution. Radical species produced during the curing, such as unreacted TPO, could also serve as potential quenching sites, an effect which has been discussed in previous work \cite{Lim2019, Frandsen2023}.

There is slight difference in PSD performance between the bulk photocured and 3D printed version of the two-dye formula, resulting from the same potential quenchers. While the bulk-cured sample was exposed to air over a short two or fewer minutes while curing, the 3D printed samples were exposed to oxygen and ambient air for over 30 minutes. Furthermore, the vat that holds the resin tilts during polymerization which agitates the resin, introducing additional oxygen. While the long per-layer curing time should limit initial radical species like TPO remaining in the geometry and reduce the quenching effect, the significant exposure to light may also be degrading the chromophores or producing new radical species within the resin and solid plastic, reducing not only LO but also separation capability. The FoM value for the 3D-printed three-dye formula is similar but slightly lower than that of the 3D-printed two-dye formula. This slight difference indicates that the tertiary dye is absorbing a small amount of emission light, but the effect is small due to the very low concentration of the dye. The separation between neutron and gamma events for the three-dye formula is just below or nearly equivalent to the 3D printed two-dye formula.

\section{Conclusion}
\label{sec:conclusion}

Additive manufacturing of plastic scintillators faces the challenge of balancing balancing geometric precision with scintillation performance. Specifically, the challenge lies in controlling the penetration depth of curing light, which typically penetrates too deeply into photopolymerizable resins, limiting the fidelity of complex 3D geometries and constraining AM designs to 2D+ approaches. Traditional strategies such as increasing concentrations of secondary fluors either fail to sufficiently attenuate cure depth or compromise the scintillation output and polymerization efficiency. This work demonstrates that by introducing a tertiary dye—decoupled from the scintillation system—print resolution can be improved without compromising light output or PSD performance. This opens the door to on-demand radiation detectors tailored to irregular environments, such as embedded detectors in medical imaging arrays, custom dosimetry tools, or neutron detection systems requiring tight tolerances. Although the resulting scintillators have a lower light output than commercial plastics like EJ-200, its unique advantage lies in its compatibility with photopolymer-based 3D printing and in applications where custom geometries, embedded features, or integrated detection architectures are required but traditional machining or casting of commercial plastics is impractical. 

In this work, plastic organic scintillators were fabricated using a commercially available LCD printer (Prusa SL1S) using non-aromatic acrylate-based resins.  Printed samples were analyzed for improvements in feature resolution due to a tertiary cure-depth-limiting dye. Three resin formula types were manufactured and tested: one based on a two-dye formula fabricated through photoinitiated bulk polymerization, one utilizing the same formula but 3D printed using the Prusa SL1S, and one 3D printed utilizing the three-dye cure depth limiting formula. Each resin had a moderate concentration of primary dopant PPO (20 wt.\% by mass) and produced LO of 50\% relative to EJ-200 and reasonable PSD capability despite containing no aromatic base. The use of tertiary dye significantly reduced excess polymerization compared to the formula containing no tertiary dye printed under identical conditions without a significant impact on either relative LO or PSD capability.

Purple coloration resulted from the interaction of PPO and TPO in the presence of near-UV photoinitiation and dissipated over 14 days at room temperature. Samples exposed to additional heat over the curing or printing period were observed to retain a more yellow appearance, indicating heat during the curing process causes damage to the chromophores within completed samples. Photoluminescence of printed samples produced emission spectra with peak emission between 420-430 nm, matching closely to the optimal wavelength for PMT absorption. The optimal concentration of tertiary dye and per-layer curing time for the smallest resolvable external and internal features was found to be 0.04 wt.\% coumarin 450 and 30 s/layer, which produced fully resolved features up to 0.7 mm for a printed layer thickness of 0.1 mm. The tertiary dye also produced improvement of small feature resolution and consistently reduced excess polymerization along the sample. The bulk cured sample showed a light output up to 48.5 \% of EJ-200 with a FoM of 1.35 over a 0.9 – 1.1 MeVee light cut. When the formula was instead 3D printed, the LO was reduced to 36 \% of EJ-200 and a FoM of 1.29, with potential for further improvement with optimized print settings. In both cases, the performance of photocured scintillators is inferior to that of industry standards due mainly to the lower aromatic fraction as well as the presence of oxygen utilized in fast AM printing. 

Future work includes further modifications to the printing setup, including using an inert environment and faster removal of heat from the solution to prevent discoloration and the presence of oxygen during printing. Another aspect of future improvement is further investigation into alternative base monomers and oligomers to improve radiation detection performance without a significant loss in feature resolution and increase in curing time. 


\section*{\raggedright \textbf{CRediT authorship contribution statement}}
Chandler Moore: Conceptualization, Methodology, Software, Validation, Investigation, Writing – original draft, Visualization.  Michael Febbraro: Conceptualization, Methodology, Investigation, Writing - Review \& Editing. Allen Wood: Conceptualization, Methodology, Investigation. Juan Manfredi: Conceptualization, Methodology, Investigation, Writing - Review \& Editing.  Daniel Rutstrom: Validation, Visualization, Writing - Review \& Editing. Andrew Decker: Conceptualization. Ryan Kemnitz: Methodology, Resources. Thomas Ruland: Conceptualization, Investigation. Brennan Hackett: Resources. Paul Hausladen: Conceptualization, Resources.

\section*{\raggedright \textbf{Declaration of competing interest}}
The authors declare that they have no known competing financial interests or personal relationships that could have appeared to influence the work reported in this paper.

\section*{\raggedright \textbf{Disclaimer}}

The views expressed in this document are those of the author and do not reflect the official policy or position of the United States Air Force, the United States Department of Defense, or the United States Government. This material is declared a work of the U.S. Government and is not subject to copyright protection in the United States. Imagery in this document is the property of the U.S. Air Force.  

\section*{\raggedright \textbf{Acknowledgments}}
This material is based upon work supported by the Department of Energy National Nuclear Security Administration through the Nuclear Science and Security Consortium under Award Number(s) DE-NA0003996 and the Defense Threat Reduction Agency under grant HDTRA1136911. This work has been supported by the U.S. Department of Energy office of Nonproliferation Research and Development (National Nuclear Security Administration, Defense Nuclear Nonproliferation R\&D Office). The authors would like to thank Szymon Adamczyk on printables.com for the CAD models used to produce the F-16 and F-104 geometry shown in this work.

\bibliographystyle{JHEP}
\bibliography{biblio.bib}

\end{document}